\documentclass[12pt,a4paper]{iopart}
\usepackage{graphicx}

\usepackage{iopams}  
\usepackage{epstopdf}
\usepackage{braket}
\usepackage{float}
\usepackage[cmyk,dvipsnames]{xcolor}

\begin{document}

\title[]{Efficient Optimization Method for Finding Minimum Energy Paths of Magnetic Transitions}

\author{A.V.~Ivanov$^{1,2}$, D. Dagbartsson$^{1}$, J. Tranchida$^{3}$, V.M. Uzdin$^{2,4}$ and  H. J\'{o}nsson$^{1,5}$}
%
\address{$^1$Science Institute and Faculty of Physical Sciences, University of Iceland, VR-III, 107 Reykjav\'{\i}k, Iceland}
%
\address{$^2$Saint Petersburg State University, 199034 Saint Petersburg, Russia} 
%
\address{$^3$Multiscale Science Dpt., Sandia National Laboratories, P.O. Box 5800, MS 1322, 87185 Albuquerque, NM, United States}
%
\address{$^4$ITMO University, 197101 Saint Petersburg, Russia}
%
\address{$^5$Department of Applied Physics, Aalto University, FI-00076 Espoo, Finland}

\ead{alxvov@gmail.com}
 
\begin{abstract}
Efficient algorithms for the calculation of minimum energy paths of magnetic transitions are implemented within the geodesic nudged elastic band (GNEB) approach.
While an objective function is not available for GNEB and a traditional line search can, therefore, not be performed, the use of limited memory Broyden-Fletcher-Goldfarb-Shanno (LBFGS) and conjugate gradient algorithms in conjunction with orthogonal spin optimization (OSO) approach is shown to greatly outperform the previously used velocity projection and dissipative Landau-Lifschitz dynamics optimization methods. 
The implementation makes use of energy weighted springs for the distribution of the discretization points along the path and this is found to improve performance
significantly.
The various methods are applied to several test problems using a Heisenberg-type Hamiltonian, extended in some cases to include Dzyaloshinskii-Moriya and exchange interactions beyond nearest neighbors. 
Minimum energy paths are found for magnetization reversals in a nano-island, collapse of skyrmions in two-dimensional layers and annihilation of a chiral bobber near the surface of a three-dimensional magnet. The LBFGS-OSO method is found to outperform the dynamics based approaches by up to a factor of 8 in some cases. 
\end{abstract}

\vspace{2pc}


\section{Introduction}

An estimation of the thermal stability of a magnetic state can be of critical importance, for example in the context of various magnetic devices. 
Each (meta)stable magnetic state corresponds to a local minimum on the energy surface characterizing the system. 
When a magnetic element is small enough and the temperature high enough,  
thermal fluctuations due to the coupling to a heat bath can induce magnetic transitions. 
The lifetime of a state at a given temperature can be estimated using harmonic transition state theory~\cite{Bessarab2012,Bessarab2013a,Bessarab2013b}. 
In addition to the local minimum characterizing the initial state, the relevant first order saddle points on the energy surface need to be found in order 
to estimate the activation energy and entropy for each transition mechanism and thereby estimate parameters in the Arrhenius rate expression.
This can be accomplished using the geodesic nudged elastic band (GNEB) method~\cite{Bessarab2015} 
for finding minimum energy paths (MEPs) between the given initial state and each one of the final states. 
A maximum in energy along an MEP corresponds to a saddle point on the energy surface. The MEP may have more than one maximum and can reveal
unexpected metastable states as intermediate minima, but for the purpose of lifetime estimation the highest maximum is the most relevant one.
Once the MEP has been estimated closely enough to reveal its shape and the location of the highest maximum has been established well enough, 
it can be efficacious to focus only on the region near the maximum and refine the MEP there~\cite{Lobanov2017} or use a saddle point search method to 
converge more thoroughly on the maximum~\cite{Muller2018}.

Previous implementations of the GNEB method have mainly used dissipative spin dynamics or 
velocity projection algorithm based on spherical 
coordinates~\cite{Bessarab2015,Vlasov2016,Moskalenko2016,Liashko2017,Ivanov2017,Bessarab2018}. 
The calculations can require substantial computational effort, especially when long range 
interactions are included or when each energy evaluation requires self-consistent-field iterations such as in density functional theory (DFT)~\cite{Kohn1999} 
or non-collinear Alexander-Anderson calculations~\cite{Bessarab2014}. 
It is, therefore, important to develop an implementation of GNEB that requires 
as few energy and energy gradient evaluations as possible to obtain convergence to an MEP. 
For rearrangements of atoms, for example in chemical reactions and diffusion events, well-documented approaches for efficient optimization of MEPs have been developed~\cite{Sheppard2008,Asgeirsson18,Asgeirsson20}, but for calculations of magnetic systems less work has been done. 
The difficulty arising for magnetic systems is an additional constraint on the length of the magnetic moments. 
As a result, conventional optimization algorithm cannot be applied without significant modification. 

A recently proposed approach to the problem of finding local minima, the orthogonal spin optimization (OSO)~\cite{Ivanov2020},
corresponds to rotations of the magnetic vectors on a sphere and can be combined with algorithms developed for unconstrained optimization. 
The rotations are performed in Euclidean space and do not require the introduction of variables confined to a sphere.
Here, we apply the OSO approach to GNEB calculations of MEPs and 
compare the performance of various optimization methods. 
The combination of OSO with the limited-memory Broyden-Fletcher-Goldfarb-Shanno (LBFGS) optimization
method is found to be most efficient in that it requires the fewest energy and gradient evaluations to reach convergence to MEPs apart from the case where the transition path levels off from an upward trajectory, and a first order saddle point is  undefined. In the latter case, the conjugate gradient algorithms appeared to be most efficient.
These methods offer a significant improvement over previously used methods for GNEB calculations. 
The algorithms have been implemented in the LAMMPS software 
which has been used in the present calculations~\cite{Plimpton1995}.

\section{Methodology}

Two types of optimization methods are tested here. The first type is based on equations of motion, either dissipative Landau-Lifshitz (dis-LL) dynamics
or effective particle dynamics using velocity projection optimization (VPO) \cite{Bessarab2015}. 
The other type involves the OSO approach~\cite{Ivanov2020} in combination with either conjugate gradient (CG) or LBFGS optimization~\cite{Nocedal2006}. 
The presentation of these methods is brief here as they have been presented in detail elsewhere
in the context of finding local minima on the energy surface. References are given to publications that provide more detail.
The GNEB method is also reviewed briefly for completeness. 

\subsection{Dissipative spin dynamics}
Minimization of the energy of magnetic systems is often carried out using 
the Landau-Lifshitz equation of motion with only the dissipation term included
\begin{equation}
    \frac{d \hat s_i}{dt} = - \alpha \hat s_i \times \hat s_i \times \vec{\omega}_i,
\end{equation}
where $\alpha$ is a damping parameter and
\begin{equation}\label{eq: omega}
\vec{\omega}_i = - \frac{1}{\hbar}
\frac{\partial E}{\partial \hat s_i} 
\end{equation}
%
with $\hbar$ being a reduced Planck constant.
This is referred to as dis-LL method and it generates a steepest descent path on the energy surface~\cite{Ivanov2020}. 
The implementation in the LAMMPS software \cite{Plimpton1995,Tranchida2018} 
is based on a simplectic integrator with adaptive time step~\cite{Tranchida2018}. At each iteration the time step is
\begin{equation}\label{eq: add_step}
    \Delta t = \frac{2\pi}{\kappa |\vec{\omega}_{\max}|},
\end{equation}
where $|\vec{\omega}_{\max}|$ is the maximum precession frequency in the system, and $\kappa$ is a parameter.  
Results are presented here for two values of $\kappa$, 10 and 20.
The smaller value gives larger time step and faster convergence but can lead to instability.
An even smaller value, $\kappa$=5, leads to convergence problems.


\subsection{Velocity projection optimisation}

Another way to generate a steepest descent path to a minimum on the energy surface is to use effective equations of motion that mimic those of a particle.
A pseudo mass is given to the spin and the orientation and velocity calculated 
using a discretization of particle equations of motion, such as the 
velocity Verlet algorithm~\cite{Andersen80}.
In order to reach the minimum, a projection is applied to the velocity associated with spin $i$ 
in such a way that at each step only the component of the velocity in the direction of the force, the negative of the 
gradient of the energy with respect to the $i$-th spin projected on the tangent space, $\vec{f}_i = - \mathcal{P}\left[ \frac{\partial E}{\partial \hat s_i}\right]$, 
is kept when the inner product of velocity and force is positive, 
while the velocity is zeroed if the inner product is negative
\begin{eqnarray}
&\beta = \sum_i \vec{f}_i \cdot \vec{v}_i / \sum_i { \vec{f}_i} \cdot \vec{f}_i \\ 
&\mathrm{if\,\,} \beta < 0 \mathrm{\,\,then\,\,set\,\,} \beta = 0\\ 
&{\vec{v}}_i \leftarrow \beta {\vec{f}}_i. 
\end{eqnarray}

In this way, the pseudo-particle can accelerate when the gradient keeps pointing in a similar direction, but the velocity is zeroed when the particle has gone past 
the minimum.  The algorithm is analogous to a variable time step algorithm tracing out the steepest descent path.
The VPO method was used in the original implementation of the GNEB method and more detail is provided in Ref.~\cite{Bessarab2015}. Here, the VPO has been used in conjunction with OSO as described in Ref.~\cite{Ivanov2020} and briefly discussed below.


\subsection{Orthogonal spin optimization}

In orthogonal spin optimization (OSO)~\cite{Ivanov2020}, the spin directions are updated according to the following iteration process
\begin{equation}
    \hat s_i \leftarrow e^{-P_i} \hat s_i 
\end{equation}
where $P_i$ is a skew-symmetric matrix corresponding to rotation towards the energy minimum. 
For a gradient descent algorithm, $P_i$ is 
\begin{equation}\label{eq: T_matrix}
P_i =
\left(\begin{array}{ccc}
0& -g_{iz}& g_{iy}\\
g_{iz}& 0& -g_{ix}\\
-g_{iy}& g_{ix}& 0
\end{array} \right),
\quad \vec g_{i} =  \hat{s}_{i} \times \frac{\partial E}{\partial \hat{s}_{i}} .
\end{equation}

In order to improve the performance of the gradient descent algorithms (dis-LL and VPO) as well as the CG algorithm (without line search) 
in GNEB calculations, the 
gradient
is scaled to make it dimensionless and the magnitude adjusted according to the adaptive time step given by Eq.~(\ref{eq: add_step}) 
\begin{equation}
    \vec{g}_i \leftarrow \vec{g}_i \frac{\Delta t}{\hbar} 
\end{equation}
while for the LBFGS algorithm a maximum-rotation parameter is used~\cite{Ivanov2020}
\begin{equation}
P_i \leftarrow P_i \frac{\theta_{max}}{\theta_{rms}} \quad {\rm if\,\,} \frac{\theta_{max}}{\theta_{rms}} < 1, 
\end{equation}
where $\theta_{rms} = \sqrt{-\sum_i\Tr\left[ P_i^2 \right]/2N}$, $N$ is the number of spins and $\theta_{max}$ is a parameter, chosen to be $J\pi/300$. 
This makes the LBFGS algorithm stable and can be particularly important in the beginning of a minimization since the line search procedure is not performed.
A description of OSO combined with CG and LBFGS optimization for finding
local energy minima is given in Ref.~\cite{Ivanov2020}.
Here, we extend that work to calculations of MEPs. The Fletcher-Reeves CG algorithm has been used.


\subsection{Geodesic nudged elastic band}

For completeness, we briefly review the GNEB approach.
It is an extension of the nudged elastic band that is frequently used in calculations of MEPs for
atomic rearrangements, such as chemical reactions and diffusion events~\cite{Mills1995,Jonsson1998}.
There, a path connecting the initial state minimum to the final state minimum is represented with a set of discrete replicas of the system, 
referred to as images. This provides a discretized representation of a path that is initially started by some interpolation between the initial 
and final states and then converges to an MEP using some iterative optimization algorithm.  
The distribution of the images, i.e.~the discretization points, 
is controlled with a spring force that acts only along the path and only between adjacent images.
GNEB takes into account that the variables represent orientation of vectors such as magnetic momenta rather than Cartesian coordinates of particles. 

The effective gradient acting on spin $i$ of image $\nu$ is
\begin{eqnarray}
    \label{eq: GNEB_grad}
    & \vec{h}_{i\nu} = \vec{h}_{i\nu}^{\perp} + \vec{h}_{i\nu}^{\|}\\
{\rm where}\ \ \ 
    &\vec{h}_{i\nu}^{\perp} = 
    \frac{\partial E}{\partial \hat s_{i\nu}} -  
    \left(\sum_{j}
        \frac{\partial E}{\partial \hat s_{j\nu}} \cdot
        \hat\mathcal{P} \hat\tau_{j\nu}
    \right)
    \hat\tau_{i\nu}\\
{\rm and} \ \ \ 
    &\vec{h}_{i\nu}^{\|} = \left[k_{\nu}D(\nu+1,\nu) - k_{\nu-1}D(\nu,\nu-1)\right] \hat \tau_{i\nu} .
\end{eqnarray}
Here, 
$\mathcal{P}$ is a projection operator, 
$\hat\mathcal{P}[\cdot] = {\mathcal{P}[\cdot]}/{\left|\mathcal{P}[\cdot]\right|}$, 
$D(\nu+1,\nu)$ the geodesic distance between images $\nu+1$ and $\nu$, 
$k_{\nu}$ a spring constant and $\hat \tau$ the unit tangent to the path estimated from adjacent higher energy image~\cite{Henkelman2000b}. 
For more detail see~Ref.\cite{Bessarab2015}.

In previous implementations of GNEB the spring constant has been taken to be the same for all pairs of images.
In the present implementation, the spring constants are chosen to be energy weighted in order to obtain better resolution of the path
in high energy regions since the main goal is to find the energy maximum along the path.

The spring constant is evaluated from
\begin{equation}\label{eq: energy-weigh. spr}
k_\nu =
\left\{ \begin{array}{ll}
    k_{max} - \Delta k \left(\frac{E_{max} - X_{\nu}}{E_{max} - E_{ref}}\right) & \mbox{if $X_\nu > E_{ref}$},\\
    k_{max} - \Delta k & \mbox{if $X_\nu \leq E_{ref}$}.\end{array}
\right.
\end{equation}
where $X_{\nu} = {\rm max}(E_{\nu}, E_{\nu+1})$ and $E_{ref}$ is the energy of the higher end-point of the path. 
The magnitude of the spring constants increases linearly towards the current maximum along the path.
This is found to significantly improve the rate of convergence 
especially in the LBFGS calculations.
Without this scaling, a larger number of images are needed to represent the path in order to obtain convergence.
The spring forces should be of similar magnitude as the 
energy gradient in order to obtain smooth convergence. Here, we chose the minimum value of $k$ to be one tenth of the exchange parameter, J, 
and it increases linearly by up to a factor of 2 according to Eq.~(\ref{eq: energy-weigh. spr}).

In GNEB calculations, the gradient vector used in the minimization algorithms
should be replaced by the effective GNEB gradient, Eq.~(\ref{eq: GNEB_grad}),
\begin{equation}
    \frac{\partial E}{\partial \hat s_{i\nu}} \leftarrow \vec{h}_{i\nu}
\end{equation}
and all scalar products in the minimization algorithms should involve summation with respect to both spin and image indices. 
In this way the optimization algorithms are applied to the whole chain of images simultaneously.

For the highest energy image, the spring forces are not included, only the effective gradient along the path.
It pulls the image towards the point of maximum energy and this 
is referred to as the climbing image~\cite{Henkelman2000}. 
The density of images can become different on the two sides of the climbing image.
In some cases another image can during the iterative optimization of the path become higher in energy 
than the climbing image. Then, the assignment a climbing image is changed to the higher energy one.
In such cases, the LBFGS algorithm is reset. 
In the calculations presented here, a climbing image has been used from the beginning of the iterative optimization of the paths.

The initial path between the two local minima is generated by a linear interpolation using Rodrigues' formula as discussed in Ref~\cite{Bessarab2015}. 
Iterative optimization of the path then brings it to the closest MEP.
In some cases, two or more MEPs connect the end-point minima. Then the
path obtained from an initial linear interpolation may not be the one corresponding to the lowest energy barrier and thereby not represent the most likely
mechanism. 
For complex systems, genetic algorithms can be used to sample the MEPs in order to find the optimal one~\cite{Maras2016}.

In some cases, the initial path is stationary in that the gradient of the energy is zero at the images even though the path does not correspond to an MEP. 
This occurs for one of the systems used here in the performance tests (system A, see below).
The initial path corresponds in this case to an energy ridge and the maximum along the path is a higher order saddle point on the energy surface.  
In order to push the initial path off the ridge so the iterative optimization can bring the images to an MEP, 
random noise is added to the spin configurations in the initial path. 
The unit vector describing the axis of rotation for spin $i$, defined as in Eq.(16) of Ref.~\cite{Bessarab2015}, is then tilted in a random way by
\begin{equation}
    \hat{k}_i \leftarrow \frac{\hat{k}_i + \vec{\xi}_i}{|\hat{k}_i + \vec{\xi}_i|},
\end{equation}
where $\vec{\xi}_i$ is a three-dimensional vector with components generated from random numbers distributed uniformly over the interval $(-\epsilon, \epsilon)$. 


\section{Simulated systems}

The energy of the systems simulated in the tests presented here is given by an extended Heisenberg Hamiltonian
\begin{equation}
    E = 
    -\sum_{i>j} 
    J_{ij} \, \hat{s}_i \cdot \hat{s}_j  +  
    \sum_{\left<i>j\right>} \vec{D}_{ij} \cdot \hat{s}_i \times \hat{s}_j  - \sum_{i}\mu_i \hat{s}_i \cdot \vec{B} - \sum_{m} K^{m} \sum_{i} (\hat{s}_i \cdot \hat{k}^{m})^2
\end{equation}
where $\hat s$ denotes unit vector in the direction of magnetic moment, $J_{ij}$ is exchange parameter for interaction between spins $i$ and $j$,  $\vec{D}_{ij}$ a Dzyaloshinskii-Moriya vector, 
$\mu_i$ the magnitude of spin $i$ in the unit of $\mu_B$ the Bohr magneton, and
$K^{m} $ an  uniaxial anisotropy associated with direction $\hat{k}^{m}$. In the last term sum with index $m$ runs over all anisotropies in the system, for example, easy axis and easy plane. $\left<i>j\right>$ denotes sum among first neighbouring spins and each interaction between pair of spins is took once.
The values of the parameters in the Hamiltonian are taken from previous simulations of these systems, as specified below.

MEPs of transitions in five different systems are calculated with the various optimization methods to compare performance. 
The systems are briefly presented below, and references given to more complete descriptions.


\subsection{System A:  Fe nanoisland on W(110) surface}

A rectangular island consisting of 30 x 10 Fe atoms sitting in a commensurate way on a W(110) substrate is simulated in the same way as in Ref.~\cite{Bessarab2013b}.
Insets of Fig. 1 show a schematic of the island. The parameters of the Hamiltonian are chosen to be
 J = 25.6 meV (only nearest neighbor interaction),
 K$_{||}$ = 1.2 meV corresponding to $\hat{k} = (0, 1, 0)^{T}$,
 K$_{\perp}$ = -0.5 meV corresponding to $\hat{k} = (0, 0, 1)^{T}$, and D=$|\vec{D}_{ij}|$=$|\vec B|$=0.
 The remagnetization takes place by formation of a temporary domain wall in the direction of the narrower island width, as shown in Fig. 1
 (the energy curve is obtained using cubic polynomial interpolation between the GNEB images
 based on both energy and gradient along the path~\cite{Bessarab2015}).
 A study of the effect of island shape and size on the mechanism and rate of remagnetization has previously been presented and 
 comparison made with experimental data~\cite{Bessarab2013b}.

\begin{figure}[t]
\begin{center}
\includegraphics[width=0.9\textwidth]{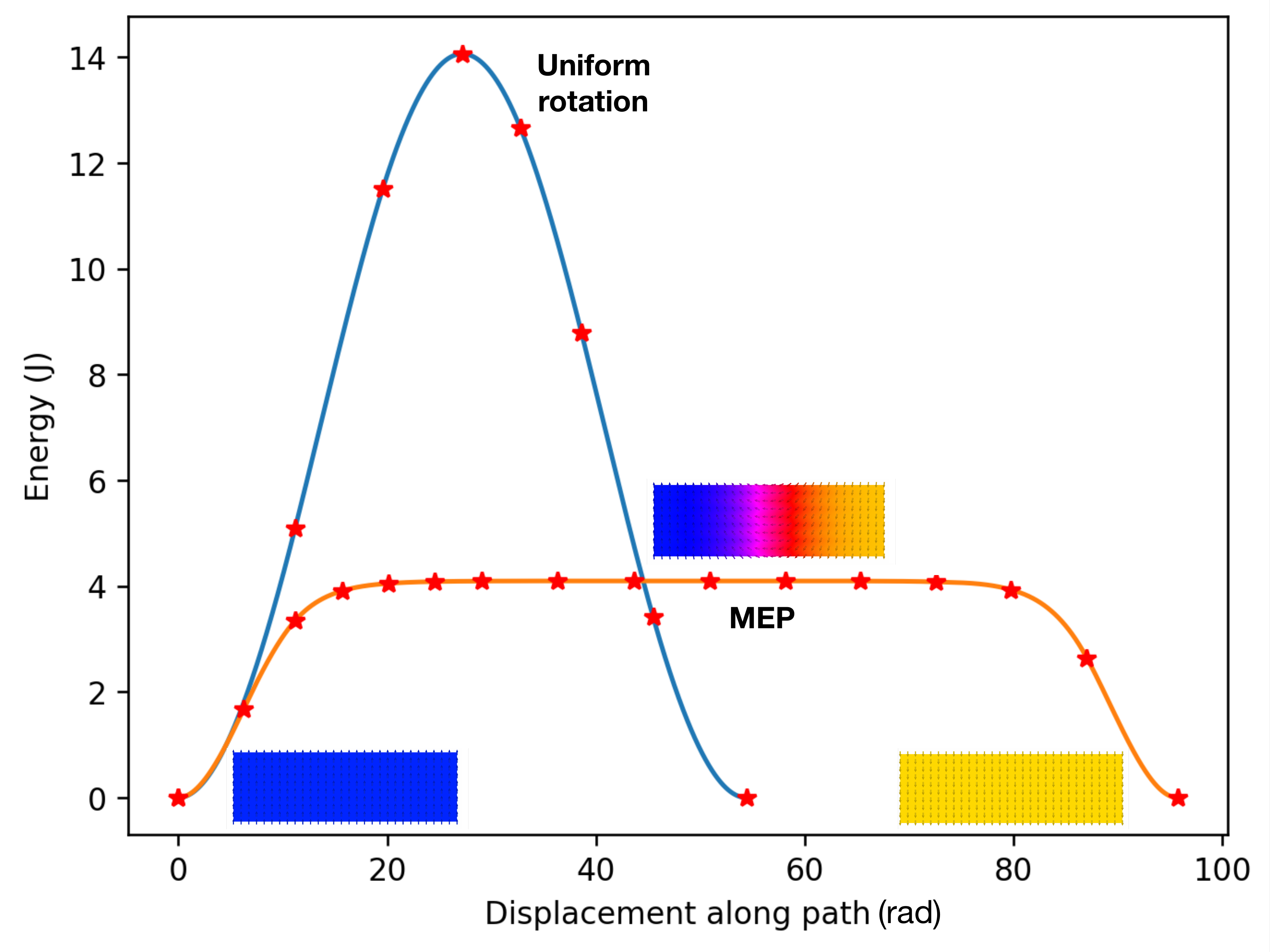}
\caption{
Magnetization reversal of an Fe nano-island on a W(110) surface. 
Blue curve:  Uniform rotation path, lying on an energy ridge going through a 
higher order saddle point, 
obtained when the initial path does not contain large enough random displacements to break the symmetry.
Orange curve: MEP corresponding to magnetization reversal starting from the right hand side of the island, 
obtained when 
random displacements with $\epsilon$=0.1 are added to the initial path.
Red stars indicate the energy and location of the images of the system in the GNEB calculations. 
The MEP is shown with more images to better illustrate the flat barrier region,
but the benchmark calculations were carried out with 8 images.
Insets show initial state, saddle point, and final state spin configurations for MEP. 
Spins lie in the plane 
of atoms and the color indicates the projection along the narrower width of the island
(color scheme:  blue corresponds to 1, yellow -1 and red 0). 
\label{fig:fig1}
}
\end{center}
\end{figure} 


\subsection{System B: Skyrmion in 2D Co overlayer on Pt}

The simulation cell contains 200 x 200 atoms subject to periodic boundary conditions in the x- and y-direction
representing a Co overlayer on a Pt(111) surface.
The values of the parameters in the Hamiltonian are  J = 29 meV  (only nearest neighbor interaction), 
D = 1.5 meV, B = 0, K = 0.293 meV, and $\hat{k} = (0, 0, 1)^{T}$
and the direction of the Dzyaloshinskii-Moriya vector is chosen to give N\'eel skyrmions.
The skyrmion state is metastable with respect to the ferromagnetic state and the calculation finds the MEP between the two states,
see Fig. 2(a).
The saddle point for collapse has only slightly higher energy than the minimum corresponding to the skyrmion in this case.
This system size and set of parameters have been used in previous studies of the stability of skyrmions in this system~\cite{Lobanov2016}.
See also Ref. \cite{Rohart2016}.


\begin{figure}[t]
\begin{center}
\includegraphics[width=0.9\textwidth]{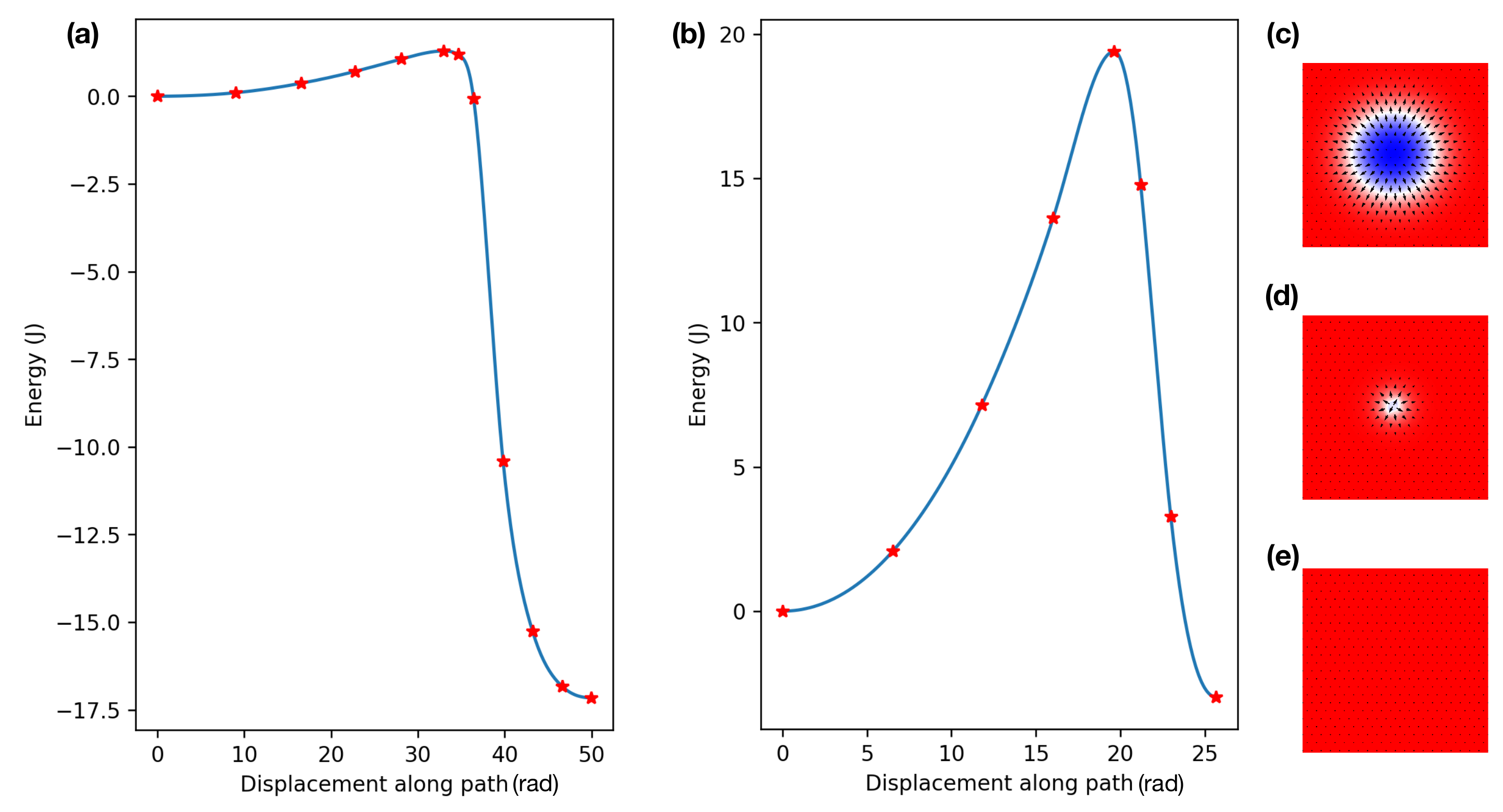}
\caption{ 
Minimum energy paths for the collapse of N\'eel skyrmions in 2D systems and formation ferromagnetic final states.
Red stars correspond to GNEB images.
(a) Co/Pt (system B). Results of a calculation using 12 images are shown, in order to better describe the path, but the benchmark calculations were carried 
out with 8 images.
(b) PdFe/Ir (system C) where
(c), (d) and (e) show the corresponding initial state minimum, saddle point, and final state minimum, respectively. 
Color indicates the out of plane spin component, blue corresponding to 1, white 0 and red -1.}
\label{fig:fig2}
\end{center}
\end{figure} 



\subsection{System C: Skyrmion in 2D PdFe/Ir with nearest neighbor exchange}

The simulation cell contains 200 x 200 atoms subject to periodic boundary conditions in the x- and y-direction
representing a face centered cubic PdFe overlayer on Ir(111).
The parameter values are  
J = 7.36 meV  (only nearest neighbor interaction), D = -2.78 meV, B = 1.5 T, $\mu_i$ = 3, K = 0.7 meV, $\hat{k} = (0, 0, 1)^{T}$. 
This system size and set of parameters have been used in previous studies of the stability of N\'eel skyrmions in this system~\cite{VonMalottki2017}.
Fig. 2(b) shows the MEP between the skyrmion minimum and the ferromagnetic minimum, and insets show the spin configurations at these 
points as well as the saddle point.
The saddle point has significantly higher energy than the skyrmion state minimum in this case.


\subsection{System D: Skyrmion in 2D PdFe/Ir with long range exchange}

This is the same as system C except that the exchange interaction includes up to 5th neighbors rather than just nearest neighbors, 
and D = -3.0 meV, B = 0.5 T.
The values of the exchange coupling constants are taken from estimates obtained using density functional theory calculations~\cite{VonMalottki2017}.
A representation of J as a function of distance between the atoms was used as input in the LAMMPS simulation software, see the Appendix. 
The long range exchange interaction together with other parameters used turns out to bring in a challenging feature 
as the gradient varies rapidly near the saddle point for skyrmion collapse, see Fig. 3.


\begin{figure}[t]
\begin{center}
\includegraphics[width=0.9\textwidth]{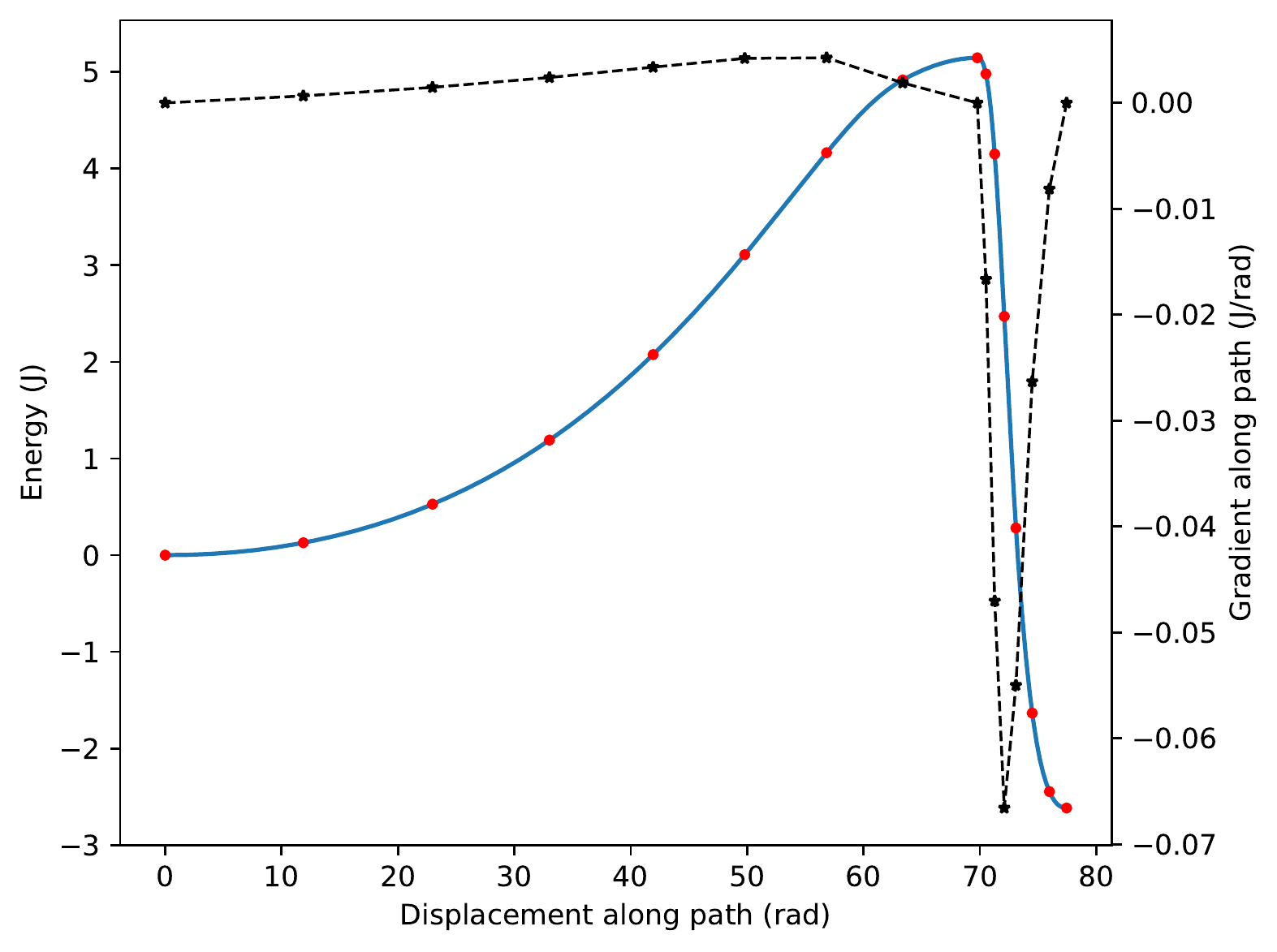}
\caption{
Collapse of a N\'eel skyrmion in a system of spins with up to fifth nearest neighbor exchange interaction (system D). 
Left axis and solid line: energy long the path. 
Right axis and dashed line: projection of the gradient vector along the path tangent.
Results of a GNEB calculation with 16 images is shown in order to better resolve the path, 
but the benchmark calculations were carried out with 8 images.
The gradient undergoes rapid change near the saddle point, so the larger number of images is needed 
to obtain convergence with the tighter tolerance of $1 \cdot10^{-6}~J$.
}
\label{fig:fig3}
\end{center}
\end{figure}


\subsection{System E: Chiral bobber in 3D system}

The simulation cell contains 30 x 30 x 30 atoms subject to periodic boundary conditions in $xy$ plane 
representing a chiral bobber located near one of the boundaries, see Fig. 4. 
The final state of the transition is the conical state. 
The parameter values are  
J = 1 meV  (only nearest neighbour interaction), D = 0.45 meV, B = 2.8 T, $\mu_i$ = 1.
No anisotropy is included for this system. The direction of the Dzyaloshinskii-Moriya vector is chosen to give Bloch type chirality.
This system size and set of parameters is the same as in reported simulation studies in Refs.~\cite{Rybakov2015,Mueller2019}.


\begin{figure}[t]
\begin{center}
\includegraphics[width=0.9\textwidth]{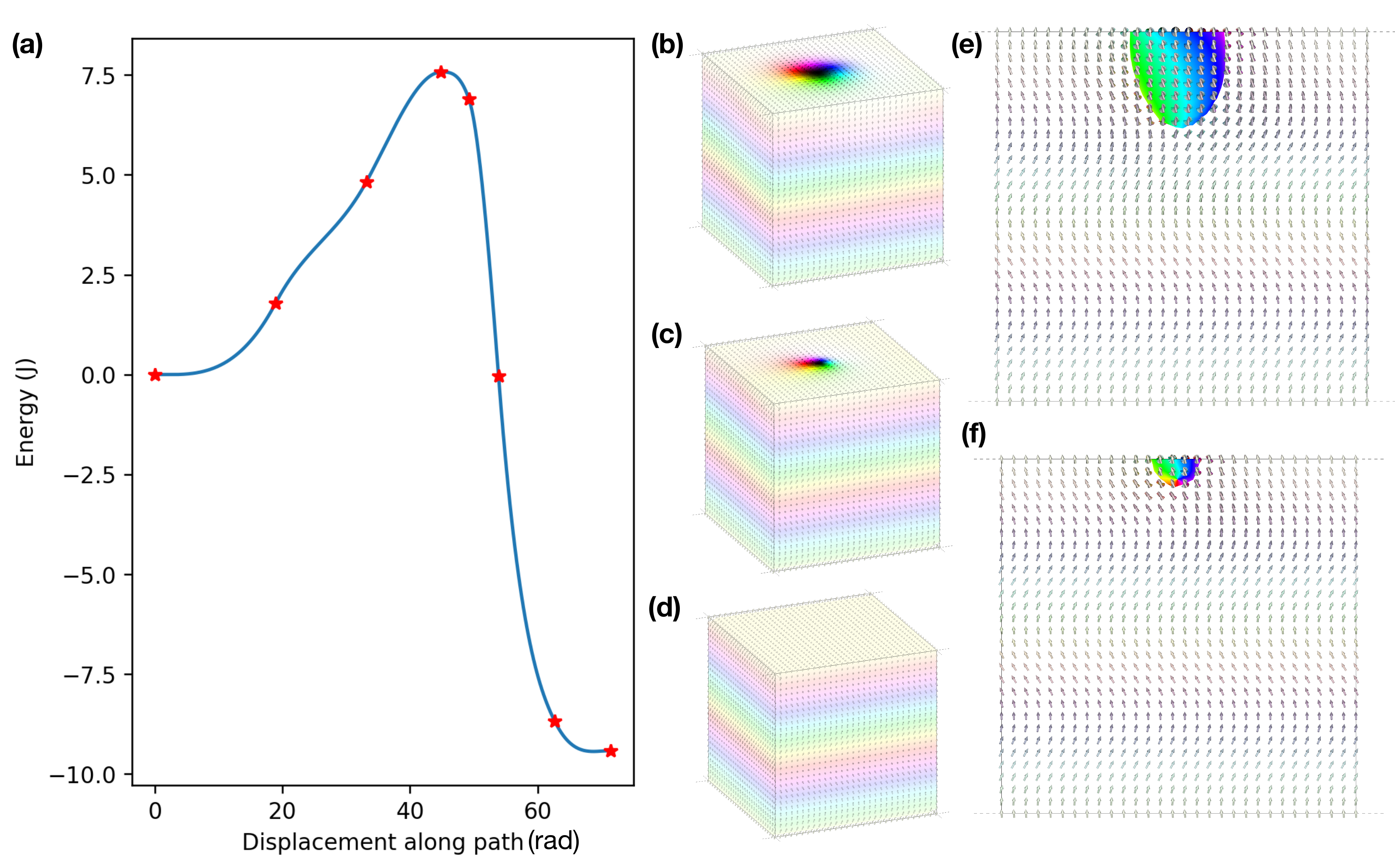}
\caption{Annihilation of a chiral bobber. 
(a) Energy along the minimum energy path. 
(b), (c), (d):  Initial state minimum, saddle point, and final state minimum, respectively. The color coding follows the  'hue, saturation, value' (HSV) model. The hue indicates the azimuthal angle while the strength of the color is determined by the polar angle. 
(e) and (f): Initial state minimum and saddle point seen from the side. 
The isosurface corresponds to spins with polar angle $\pi/2$ while the color represents the azimuthal vector.  
The inset graphics were generated using the Spirit software~\cite{Mueller2019}. 
}
\label{fig:fig4}
\end{center}
\end{figure} 

\section{Results} 
\label{sec: results}

The results of the performance tests are summarized in Table 1. The number of iterations needed to reach convergence using a total of 8 images 
(i.e. 6 movable images) to represent the GNEB path is reported. 
For all four optimization methods, there is one energy and gradient evaluation per movable image for each iteration. 
The OSO-LBFGS method outperforms the other methods for problem B-E, especially those based on equations of motion, namely the dis-LL and VPO methods. 
In some cases the OSO-LBFGS method is an order of magnitude more efficient than the frequently used dis-LL method.
For problem A, however, the OSO-CG algorithm requires fewer iterations than OSO-LBFGS.

The value of the $\kappa$ parameter in the variable time step expression Eq. (3) has significant effect on the number of iterations needed to obtain 
convergence with the OSO-GC and dis-LL methods. If the value of $\kappa$ is increased from 10 to 20, the time steps are half as smaller and the 
number of iterations correspondingly double. This has, however, only a minor effect on the number of iterations needed for the VPO method. 
There, the system accelerates when the gradient keeps pointing in a similar direction, as will occur more often when a smaller time step is used.
As a result, the algorithm effectively adjusts the magnitude of the displacement.   For system A, the number of iterations needed for the VPO algorithm 
even decreases when a larger value of $\kappa$ is used. 

The energy weighted spring forces make it possible to reach convergence with fewer images as the density of images becomes higher 
near the saddle point. A certain density is needed in order for the tangent estimate in the discrete GNEB representation of the path to be accurate enough. 
For example, in calculations with OSO-LBFSG for system C,
convergence is not reached with equal spring constants and 6 movable images when the tolerance is set to $1 \cdot10^{-6}~J$. 
However, when the number of movable images is increased to 12, convergence can be reached.  
At the same time the number of iterations required increases to 267 (from 101).
The use of energy weighted spring constants, therefore, decreases the computational effort significantly - by a factor of four in this case.

\begin{table}[!ht]
\begin{center}
\label{tab:table2}
\caption{  
Number of iterations required in order to converge on minimum energy path for the five systems studied, A-E.  
The GNEB calculation is considered to be converged when  the effective torques with GNEB gradient Eq.~(\ref{eq: GNEB_grad}) with respect to 
each one
of the spins in 
all
the images
has a magnitude below a tolerance of $1 \cdot10^{-6}~J$,
except for system D where a tolerance of $5 \cdot 10^{-6}~J$ is used, as none of the optimization methods give convergence for the smaller value 
when 8 images (6 movable) are used. 
This is because of the sharp variations in the gradient near the saddle point as the skyrmion collapses when long range exchange interactions 
are included in the Hamiltonian. 
The results for OSO-CG, VPO and dis-LL were obtained with $\kappa = 10$ in Eq. (3). Results obtained with $\kappa = 20$ are shown in brackets.
}
\vskip 0.3 true cm
\begin{tabular}{ c c c c c}
\hline
\hline
 System & OSO-LBFGS & OSO-CG & VPO & dis-LL\\
\hline
A & 814 & 505 (976) & 2425 (1629) & 1408 (2813)\\
 B & 102 & 313 (623) & 590 (685) & 888 (1776)\\ 
C & 101 & 164 (323) & 610 (643) & 455 (911) \\
D & 89 & 253 (518)& 676 (747) & 725 (1448)\\
E & 180 & 455 (893) & 968 (1086)& 1299 (2610)\\
\hline  
\end{tabular}
\end{center}
\end{table}

A special issue arises for the magnetization reversal of the Fe nano-island, system A.
The initial path is a stationary path, i.e. with zero energy gradient, because of symmetry and it lies along an energy ridge going through a higher order saddle point on the energy surface.
It is necessary to break the symmetry in some way for the iterative GNEB calculations to converge to an MEP.
The optimal reversal mechanism corresponds to a temporary domain wall in the direction of the narrower width of the island 
and starting from either one of the two ends of the island.
As the domain wall propagates through the island, the energy is practically constant, as can be seen from Fig. 1.
Another MEP corresponds to a temporary domain wall in the direction of the wider width of the island but it corresponds to higher energy barrier.
By adding large enough random rotations of the spins in the initial path as described in Sec. 2, Eq. (16), 
the iterative optimization of the path converges to either one of the two optimal and symmetrically equivalent MEPs.

Table 2 gives the number of iterations needed for several different calculations of this system.
Even when no random rotations are added and the path remains on the energy ridge, it takes a few iterations to converge because of the climbing image. 
The rise in energy along the path is large, 14.062 J (see Fig. 1).
The initial path contains an even number of images so no image is at the maximum. 
With an odd number of images, the middle one would be sitting at the saddle point.  
When small random displacements are added to the initial path with $\epsilon$=0.01, the iterative optimizations still converge on the energy ridge.
Points on the ridge are at a maximum along the direction of some of the vibrational modes orthogonal to the path, but along all other modes they are at a minimum.
With larger random displacements, $\epsilon$=0.1, the symmetry is broken sufficiently strongly
for the optimization to converge on an MEP where the rise in energy is only 4.097 J, corresponding to the activation energy for thermal transitions. 
The uniform rotation is, therefore, far from being the preferred mechanism for thermally activated magnetization reversals.
For small enough islands with less than 14 atoms along each side, the uniform rotation mechanism becomes preferred~\cite{Bessarab2013b}. 

The number of iterations needed for convergence is different for different sets of random numbers, 
generated from a different random number seed, as can be seen from the last two lines in Table 2.
This is to be expected since the initial paths are different.


\begin{table}[!ht]
\begin{center}
\label{tab:table3} 
\caption{ 
Number of iterations needed for convergence in the GNEB calculations
of remagnetization in the 30 x 10 Fe island on a W(110) surface (system A).
The tolerance for convergence is $1 \cdot10^{-6}~J$.
Results are given for various values of the magnitude of the random rotations of spins in the initial path, $\epsilon$ in Eq. (16).
For the largest value, $\epsilon$=0.1, results of calculations using two different random seeds are given 
(one being the same as in Table 1). 
Those calculations both converge on the minimum energy path with a maximum of 4.097 J corresponding
to the activation energy, E$_a$.
Calculations started from paths with $\epsilon$=0 or 0.01 converge on the uniform rotation path with a 
maximum of 14.062 J.  
The results for OSO-CG, VPO and dis-LL were obtained with $\kappa = 10$ in Eq. (3). Results obtained with $\kappa = 20$ are shown in brackets.
}
\vskip 0.3 true cm
\begin{tabular}{ c c c c c c}
\hline
\hline		
 $\epsilon$ &  
 {OSO-LBFGS}  & 
 {OSO-CG} & 
 {VPO} & 
 {dis-LL} \\

\hline		
\hline
$0$  
& 90 
& 84 (160)
& 278 (426)
& 233 (466)
\\
$0.01$  
& 76 
& 92 (173)  
& 274 (402)
& 235 (470)
\\
$0.1$  
&1044 
& 436 (850)
&1847 (1697)
& 1230 (2459)
\\
%
$0.1$  
& 814 
& 505 (976)
& 2425 (1629)
& 1408 (2813)
\\
\hline  
\end{tabular}
\end{center}
\end{table}


\section{Conclusion}

The results of the calculations presented here show that the OSO-LBFGS optimization method gives the best performance of the four 
methods tested here, apart from 
system A
where the transition mechanism involves formation and then propagation of a domain wall 
through an island.
There, the OSO-CG method turned out to require fewer iterations when the smaller value of $\kappa$ is used.
The number of energy and gradient evaluations needed for convergence 
of the OSO-LBFGS method 
is in some cases an order of magnitude
smaller than for the commonly used dis-LL method. This performance difference is particularly important when calculations are carried out
using sophisticated models of the magnetic system, such as long range interactions 
(such as dipole-dipole interaction) and self-consistent field methods.
In such cases the evaluation of the energy and gradients of the energy account for most of the computational effort. 

The present implementation of the GNEB method makes use of energy weighted spring constants. While this was proposed two decades ago
in the context of atomic rearrangements~\cite{Henkelman2000}, it has not become commonly used and has not been implemented previously
for magnetic systems. It is found to enhance the performance of the optimization methods, especially the OSO-LBFGS method. 
Partly this is the result of an improved tangent estimate when the density of images is higher near the climbing image.  
The magnitude of the spring forces can also influence the convergence rate through the approximations of the Hessian in the LBFGS method.
There, the spring force and energy gradient are coupled together. 

In addition to the information about performance of the various optimization methods, 
the calculations present here reveal an interesting effect that 
be present
when the long range exchange interaction is included in the Hamiltonian.
Near the saddle point for annihilation of a skyrmion, the gradient of the energy 
changes abruptly, as shown in Fig. 4.
As a result, the climbing image  
may not converge on the saddle point unless the resolution of the path is high enough and the displacement steps in the optimization algorithm are small enough.
The density of images near the saddle point 
needs to be high enough to resolve the oscillations in the gradient in order for the GNEB calculation to converge.


\section{Acknowledgment}

We thank Gideon M\"uller, Moritz Sallerman and Pavel Bessarab for helpful discussions.
This work was funded by the Icelandic Research Fund, the University of Iceland Research Fund, 
Russian Foundation of Basic Research (grant RFBR 18-02-00267 A).
AVI is supported by a doctoral fellowship from the University of Iceland. 
Sandia National Laboratories is a multimission laboratory managed and operated by National Technology \& Engineering Solutions of Sandia, LLC (a wholly owned subsidiary of Honeywell International Inc.) for the U.S. Department of Energy’s National Nuclear Security Administration under contract DE-NA0003525.
The calculations were carried out at the Icelandic High Performance Computing facility.


%

\appendix
\section{Long range exchange interaction in system D}

\begin{figure}[b]
\begin{center}
\includegraphics[width=0.6\textwidth]{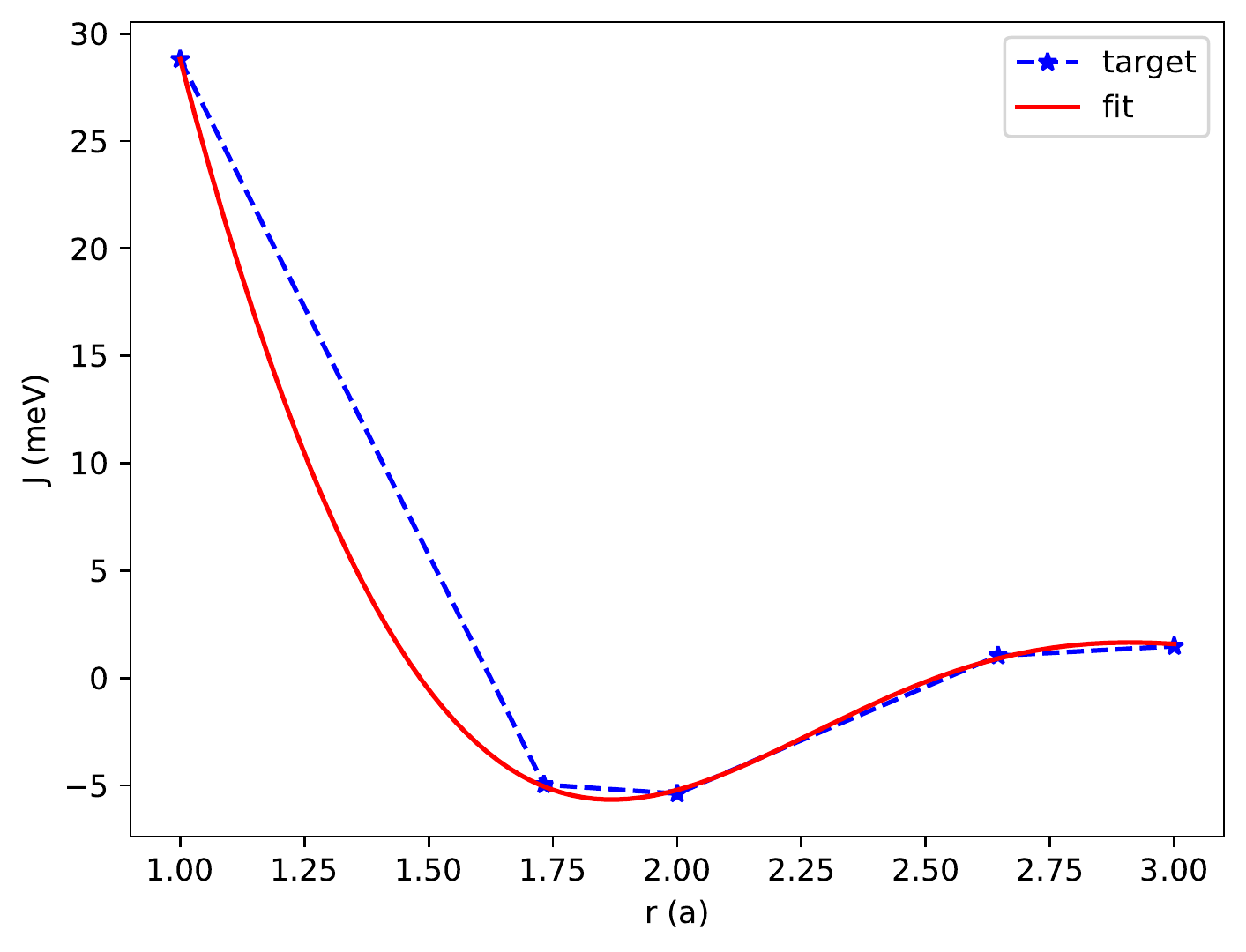}
\caption{An interpolation of the values of the exchange coefficient, J,  calculated using DFT and presented in Ref.~\cite{VonMalottki2017} to provide the 
exchange coefficient as a continuous function of the distance between the two atoms.}
\label{fig: appfig1}
\end{center}
\end{figure} 
%

In system D representing face centered cubic PdFe overlayer on Ir(111) the exchange interaction includes up to 5th neighbors. 
The values of the exchange coefficients have been estimated using DFT calculations and are presented in Ref.~\cite{VonMalottki2017}.
But, in the input for the LAMMPS software~\cite{Plimpton1995} the exchange interaction needs to be specified as a 
continuous function of the distance between the atoms.
The DFT values are, therefore, interpolated using the functional form
\begin{equation}
    J(r) = 2 a \frac{\sin(br)}{r^2} +
           2 c \frac{\sin^3(dr)}{r^2}
\end{equation}
with parameter values
\begin{eqnarray}
    a &= 11.5852\\
    b &= 2.4973\\
    c &= 8.0713\\
    d &= 1.3404 .
\end{eqnarray}
From this interpolation function, the following set of J values is obtained (the target DFT values from Ref. are shown in brackets)
\begin{eqnarray}
J_1 & = 28.81\quad(28.8) \\
J_2 & = -5.05\quad(-4.96)\\
J_3 & = -5.21\quad(-5.38)\\
J_4 & = 0.91\quad(1.04)\\
J_5 & = 1.59\quad(1.48)
\end{eqnarray}

\vfill
\eject

\vskip 1 true cm



\end{document}